\documentclass[iop]{emulateapj}
\usepackage{url}
\usepackage{graphicx}
\usepackage{xcolor}
\usepackage{soul} 

\usepackage{color}
\usepackage{amsmath}
\usepackage{setspace}
\usepackage{threeparttable}
\usepackage{epstopdf}
\usepackage{graphics}
\usepackage{subfigure}
\usepackage{url}             
\usepackage{rotating}
\usepackage{lipsum}
\usepackage{float}
\usepackage{wrapfig}
\usepackage{longtable}
\usepackage{natbib}
\usepackage{hyperref}

\usepackage{enumitem}

\graphicspath{{./}{figures/}}

\begin{document}
\nocite{*}
\title{ A QPO in Mkn 421 from Archival $\it{RXTE}$ Data}

\author{Evan Smith, Lani Oramas, Eric Perlman}
\affiliation{Aerospace, Physics and Space Sciences Department, Florida Institute of Technology, 150 West University Blvd., Melbourne, FL. 32901, USA }



\begin{abstract} 

We report a 325($-7, +8$) day quasi-periodic oscillation
(QPO) in the X-ray emission of the blazar Mkn 421, based on data obtained with the  $\it{Rossi~X}$-$\it{ray~Timing~Explorer~(RXTE)}$ satellite.
The QPO is seen prominently in the ASM data (at least 15 cycles), due to the fact that it has had near-continuous sampling for more than a decade.  The PCA data, where the sampling is not uniform and shows many large gaps, provide supporting evidence at lower significance. 
QPOs are an important observable in accretion disks, can be modulated by various orbital timescales, and may be generated by a number of mechanisms.  They have been studied extensively in X-ray binaries, and should be present in active galactic nuclei (AGN) if they are governed by a common set of physical principles.  In jetted sources, QPOs can probe jet-disk interactions or helical oscillations.
This QPO previously has been claimed intermittently in X-ray, radio and gamma-ray data, but the continuous, 15-year extent (1996-2011) of the ASM observations (in which Mkn 421 is the brightest AGN observed) provides a unique window. The QPO appears present for nearly the entire extent of the ASM observations.  We explore various physical origins and modulating mechanisms, particularly interpretations of the QPO as a result of disk-jet interactions, either due to an accretion disk limit cycle, jet instabilities or helical motions.  Limit-cycle related oscillations would not interact with either Keplerian or Lense-Thirring modulated oscillations, however those associated with jet instabilities or helical motions in the jet would likely be modulated by Lense-Thirring precession.

\end{abstract}


\section{Introduction}

 Quasi-Periodic Oscillations (QPOs) in accretion disks have been thoroughly studied in both neutron star (NS) and black hole X-ray binaries (BHXRB). High-frequency (HF) QPOs in galactic sources are thought to probe the motion of matter under strong gravity near the surface of the NS or event horizon of the BH. Models of QPOs include Keplerian orbital motion of matter in the disk, central compact object spin, general relativistic effects such as Lense-Thirring precession, or beat frequencies between two of the previous mechanisms. The history of these models can be found in various reviews such as \cite{1989ARA&A..27..517V, 2000ARA&A..38..717V} and \cite{2019NewAR..8501524I}.

$\it{RXTE}$ observations of microquasars found QPOs at nearly identical frequencies: 67 Hz for GRS 1915+105 and 66 Hz for IGR J17091-3624. Similarities in the complex light curves of these two sources prompted investigation of whether the same physical mechanisms cause these behaviors \citep{2011ApJ...742L..17A}. A cardinal question in astrophysics is whether galactic black holes in BHXRB and supermassive black holes (SMBHs) in AGN are governed by shared physical processes. \cite{2003MNRAS.345.1057M} investigated the disk-jet connection, yielding tight correlations between BH mass, radio and X-ray luminosity over many orders of magnitude, thus supporting the theory of scale invariance. If there is a commonality in the underlying physics of disk-accretion and jet-launching in BHXRB and AGN, observational evidence of timing features would appear on much different time scales, often compared using the mass ratio.


More recent studies into microquasars have given rise to new theories behind QPOs and the newly named quasi-periodic eruptions (QPEs) \citep{2020A&A...636L...2G,  2019Natur.573..381M}, expanding the list of possible common mechanisms.  The same concept of shared processes applies to galactic BHs in BHXRB and SMBHs in AGN. Similarly, \cite{2011MNRAS.414.2186J}’s expansion upon \cite{1992ApJ...385...94C,1993ApJ...419..318C,1998ApJ...494..366C}’s work has led to the accretion disk limit-cycle being considered in both AGN and BHXRB.

The widely known Schwarzschild radius is given by:

\begin{equation}
    R_S \equiv \frac{2GM}{c^2}
    \label{schwarschild radius}
\end{equation}

For a non-rotating, or Schwarzschild, BH, the innermost stable circular orbit (ISCO) for material in the accretion disk has a radius of  $R_{ISCO} \equiv 3 R_{S}$. For a rotating, or Kerr, BH the radius of the ISCO for a maximally-rotating BH in both the co-rotating and counter-rotating cases is constrained by $0.5 R_{S} \leq R_{ISCO} \leq 4.5 R_{S}$ \citep{2015PhRvD..91l4030J}. 

 Early searches for QPOs focused heavily on Seyfert galaxies, including studies conducted by {\cite{2008Natur.455..369G}; \cite{2010MNRAS.403....9M, 2014MNRAS.445L..16A, 2015MNRAS.449..467A, 2018ApJ...860L..10S, 2014ApJ...795....2E, 2018NatCo...9.4599Z, 2020ApJS..250....1T} and \cite{2020ApJ...902...65S}. Successes in these campaigns paved the way for similar searches in blazars. Much of the variable emission in blazars, particularly BL Lacertae objects, comes from non-thermal processes, particularly within the relativistic jet which is seen at very small angles to the line of sight and therefore Doppler boosted (see, e.g., \cite{Giommi_2021}).  Nevertheless, recent work has begun to suggest that the violent, variable emission in these objects can be traced to turbulence in the jets launched by their accretion flows \citep{2017ApJ...843...81O}, accentuated by Doppler boosting.  This has become the leading paradigm to explain some of the long-term variability of blazars \citep{galaxies7010035}.  Disk behavior can produce some variability in the optical and X-ray. However, jet variability must dominate in the radio and gamma-rays. The link between disk behavior and jet variability is thus quite interesting. 
 
 In the radio, QPOs have been detected in the variability behavior of the FSRQ J1359+4011 \citep{2013MNRAS.436L.114K}. In the gamma-rays, data taken with the \textit{Fermi Gamma-ray Space Telescope} have also proven fruitful in the search for QPOs. These detections include a 34.5-day QPO detected in PKS 2247$-$131 and the 612-day QPO in PKS 2155$-$304 \citep{2018NatCo...9.4599Z, 2020ApJS..250....1T}.  Indeed, in a recent campaign on CTA 102, the gamma-ray and optical light curves were both observed to have QPOs with a dominant period of 7.6 days that lasted for about 8 cycles, a behavior attributed in that paper to enhanced emission from a helical jet \citep{2020A&A...642A.129S}. In Mkn 421, two recent works utilizing \textit{Fermi LAT} data, \citep{2020ApJ...905..160B,2022arXiv220413051R}, claimed QPOs at $\sim$280 days and 308 $\pm$ 99 days respectively, although the latter authors estimated a modest 2.9$\sigma$ significance. In contrast \cite{2017MNRAS.472.3789C}, \cite{2020ApJS..250....1T}, and \cite{2017A&A...600A.132S} did not find a QPO in \textit{Fermi LAT} data of Mkn 421 despite their data including the same time frame. 
 
 However, claims of quasi-periodic behavior in blazars in the X-ray and optical have been less common.
 In Mkn 421, a recent power density spectrum (PDS) analysis of its long-term X-ray variability using data from \textit{AstroSAT} came to the conclusion that the X-ray variability was due to changes in the accretion disk translating into the jet \citep{Chatterjee_2018}, due to a break in the PDS.  That work did not search for QPOs,  but  \cite{2016PASP..128g4101L} did a combined analysis of the \textit{Swift, Fermi LAT} and \textit{Owen's Valley Radio Observatory (OVRO)} data between 2008-2015.  Those authors noted $\sim$300 day QPOs in all three light curves, but offered little information on their overall significance or persistence. 
 In the optical, the Kepler observations of the blazar W2R1926+42 \citep{2013ApJ...766...16E} had significant PDS breaks, but no QPOs. On the other hand, there is the notable case of the blazar OJ 287, where the optical light curve shows evidence of a 12-year oscillation \citep{2016ApJ...832...47B}, believed to be due to a binary SMBH \citep{2012MNRAS.427...77V}.


In addition to QPOs, subsequent X-ray flares can appear in light curves. This new phenomenon is known as quasi-periodic eruptions (QPEs). \cite{2020A&A...636L...2G} and \cite{2019Natur.573..381M} define QPEs as rapid and recurrent X-ray bursts above a stable level. Theorized causes of QPEs include radiation pressure instability, the orbital motion of a second body, tidal disruption events, and accretion disk emissions "leaking" through windows. The first confirmed QPEs are in the light curve of GSN 069, leading to a search for similar AGN. The search led to RXJ1301.9+2747, chosen based on a shared set of qualities with GSN 069. While Mkn 421 does not share these characteristics, it is possible that some massive AGN may also exhibit partial QPEs \citep{2019Natur.573..381M}. Six similar flares are seen in Figure \ref{fig3:dbdLC}, though they are not confirmed to be QPEs.

\section{Observations}
To search for low-frequency (LF) QPOs, we investigated archival data from the $\it{Rossi~X}$-$\it{ray~Timing~Explorer}$ ($\it{RXTE}$) satellite collected between 1996 and 2011. AGN were targets for monitoring with \textit{RXTE} both with the Proportional Counter Array (PCA) and All-Sky Monitor (ASM). This analysis uses data from the $\it{RXTE}$ PCA \citep{2006ApJS..163..401J} and ASM since Mkn 421 is nearly an order of magnitude brighter than any other AGN in the 2$-$10 keV bandpass shared by the PCA and ASM.

\subsection{ASM Observations}

The All-Sky Monitor (ASM) contains three rotatable coded aperture cameras called scanning shadow cameras (SSCs) \citep{1996ApJ...469L..33L}. Each SSC has position-sensitive proportional counters (PSPCs) that view the sky through slits and measure the displacements and strength of shadow patterns cast by X-ray sources in the field of view. The carbon-coated quartz anodes in the PSPCs then create a pair of pulse heights used to determine event energy and positions via the charge division technique. This technique can resolve the position (the coordinates perpendicular to the slits) within 0.2 to 0.5 mm. For each anode, the counts are binned as a function of position in three energy bands and histograms are accumulated in "dwells". Dwells are the 90 seconds during which the spacecraft maintains a fixed attitude and the orientation of each SSC is fixed with respect to the sky because the ASM's rotation drive is not active \citep{1996ApJ...469L..33L}. These position histograms are then used to calculate intensities using a least-squares fit. The fitted intensities and their uncertainties are then used to calibrate the SSCs and ensure that all three SSCs give the same results. 

The ASM scans 80\% of the sky every $\it{RXTE}$ orbit. The light curves obtained by the ASM can be used in two formats: Dwell-by-Dwell or 1-Day Average. The Dwell-by-Dwell format includes every ASM dwell for the object while the 1-Day average format gives the average of all dwells on the given date (determined by MJD). 

 For most AGN, the RXTE ASM data were of limited use since the fluxes were below the sensitivity level of the ASM. However, Mkn 421 is by far the brightest AGN in the RXTE band-pass, so it was easily detectable. Various groups \citep{2009ApJ...698.1207C, 2009ApJ...696.2170R} have investigated other AGN of lower flux by substantially binning the data.
 
\subsection{PCA Observations}

A total of 1182 pointed $\it{RXTE}$ observations were made of the BL Lac object, Mkn 421. 
AGN monitoring was regularly performed for many sources, and reduced 3-color light curves have been prepared and archived at the University of California, San Diego (UCSD) \citep{2011ApJS..193....3R,2013ApJ...772..114R}. 

$\it{RXTE}$ conducted observations on a total of 153 AGN over its 16-year lifespan. UCSD created the $\it{RXTE}$ AGN archive by collecting all the data gathered by different groups of proposers for various scientific aims, regardless of sampling strategy. Of the AGN in the UCSD database, we found 76 with enough separate observations and appropriate spacing to support the search for low-frequency QPOs. 

Details of the data reduction for the PCA data can be found in the description of the UCSD archive. A standard screening procedure was to exclude data within 20 minutes of $\it{RXTE}$ passage through the South Atlantic Anomaly (SAA). Observations made when RXTE was near the SAA could be expected to stand out from other observations since, even if the background model is very good, observations with higher background count rate will also have a larger uncertainty in the background count rate. 

The typical length of an observation is around 2000 seconds and is set by the observable times for the object within an orbit. Each observation represents a single data point in the light curve, regardless of its length. The mean flux over the entire observation is the archived value, and the time tag is the observation midpoint. For each AGN observation, the $\it{RXTE}$ AGN archive at UCSD provides the mean count rate and its uncertainty in the 2$-$10 keV spectral range, and for three sub-bands (2$-$4 keV, 4$-$7 keV, and 7$-$10 keV). Other preliminary results using the UCSD archive, both positive and negative, have been reported \citep{2019HEAD...1710631S,2020ApJ...902...65S,2020AAS...23621201H,2020AAS...23621204R,2021AAS...23752904O}.


\section{Data Analysis}
\subsection{Quality Assurance}
A useful quality assurance technique is to compute the window transform, or Lomb-Scargle periodogram (LSP) of the window function. This is done using the time series from the actual observations, but with constant flux. This can help rule out false peaks resulting from the data distribution \citep{2018ApJS..236...16V}. We find it useful to plot the window LSP together with the data LSP, allowing for easier visual comparison. Applying this technique to our analysis of Mkn 421 helps establish that the 320 day period does not result from the time spacing of the observations.

To formally establish a False Alarm Probability (FAP), we used the method of Horne $\&$ Baliunas \citep{1986ApJ...302..757H} to estimate the uncertainties in the QPO frequency, as detailed in \cite{2020ApJ...902...65S}.

\subsection{RXTE Data Analysis}
To analyze the power spectrum of Mkn 421, the $\it{RXTE}$ AGN light curve files were run through a program to determine the LSP. The input data is loaded into three arrays:
\setlist{nolistsep} 
\begin{enumerate}[noitemsep]
\item x (the MJD midpoint of the observation),
\item y (the flux in the selected band),
\item dy (the uncertainty in the flux).
\end{enumerate}
The program outputs the light curve and the LSP. In Figures 1-8, discussed below, we show the light curves and LSP for the RXTE observations.  We split our analysis into the ASM data, discussed in Section 3.2.1, and the PCA data, discussed in Section 3.2.2.

\subsubsection{ASM Analysis}

The ASM 1-Day Averages give a light curve spanning 5837 days from MJD 50091$-$55928 (1996 January 9 $-$ 2011 December 29). This is binning performed by the ASM team. 

We filtered flux values based on their signal-to-noise ratio to eliminate outliers with higher errors and filter out negative flux values seen in the raw data. A limiting signal-to-noise ratio of $S/N \geq 0.6$ accomplished both goals. However, the same spectral features are seen with and without the filtering process.

There were originally 5837 data points with a mean spacing of 1 day. After filtering, we were left with a total of 3935 data points with a mean spacing of 1.47 days. Our time frame of interest, MJD 50152$-$55924, contained 3928 data points (Figure \ref{fig1:ASM_LC}). The LSP for these data (Figure \ref{fig2:1da_PS}), with window LSP in red, shows a strong peak at 0.001708 $\mu$Hz, corresponding to a period of 650 days. This 37-$\sigma$ detection has a FAP $\sim 10^{-12}$. The peak to the immediate left is almost as high and at the 3:1 resonance with a frequency of 0.00119 $\mu$Hz, corresponding to a period of 975 days. A third strong peak is seen to right of the 0.001804 $\mu$Hz peak at 0.03552 $\mu$Hz, corresponding to a period of 325 days. This 32-$\sigma$ detection has a FAP of $\sim10^{-10}$. 

\begin{figure*}[b]
    \centering
    \includegraphics[scale=3, width=0.49\linewidth]{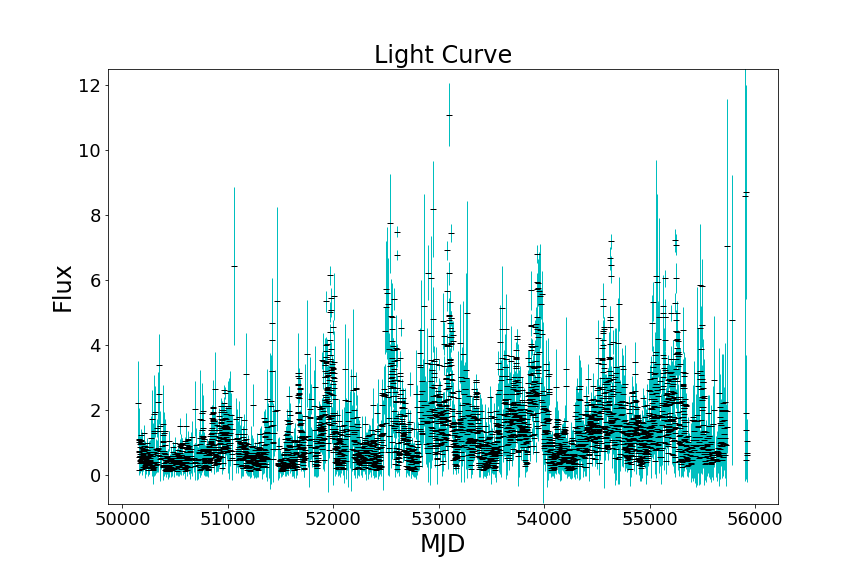}
%
    \includegraphics[scale=3, width=0.45\linewidth]{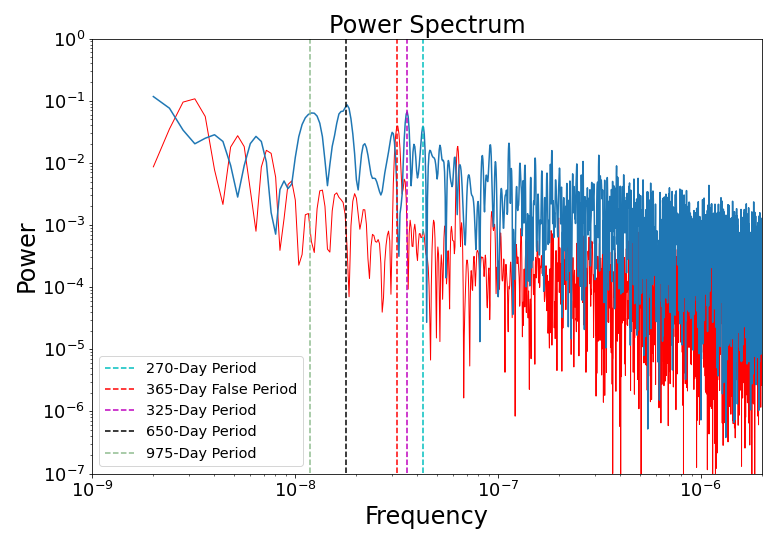}
   \caption{(left) The $\it{RXTE}$ ASM light curve for Mkn 421, 1-day averages, where the flux is in $erg \cdot cm^{-2} s^{-1} $. The data have been filtered as described in \S 3.2.1.  The essentially continuous, 15-year-long duration of the light curve is notable.}  
   \label{fig1:ASM_LC}
    \caption{(right) Mkn 421 Lomb-Scargle power spectrum (blue) for the ASM 1-day averages. The window power spectrum is shown in red.  Three strong peaks are seen: one at $0.036 \mu$Hz, corresponding to a 325-day period, another at 0.001781 $\mu$Hz (650 days), and a third at 0.00119$\mu$Hz (975 days).  These represent the primary, 2:1 and 3:1 resonances of a 325-day QPO (\S 3.2.1). The false peak at 365 days appears only in the window power spectrum.}
    \label{fig2:1da_PS}
\end{figure*}

\begin{figure*}[b]
    \centering
    \includegraphics[scale=3, width=0.49\linewidth]{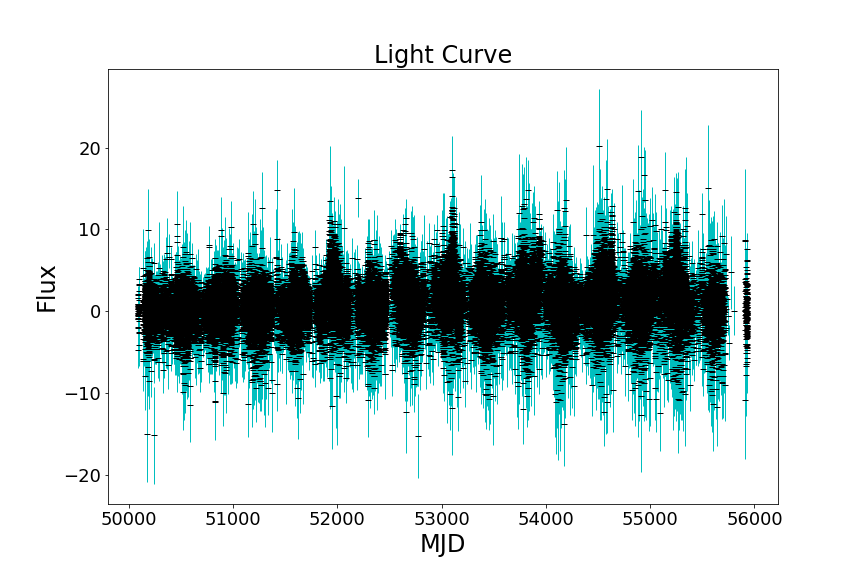}
%
    \includegraphics[scale=3, width=0.49\linewidth]{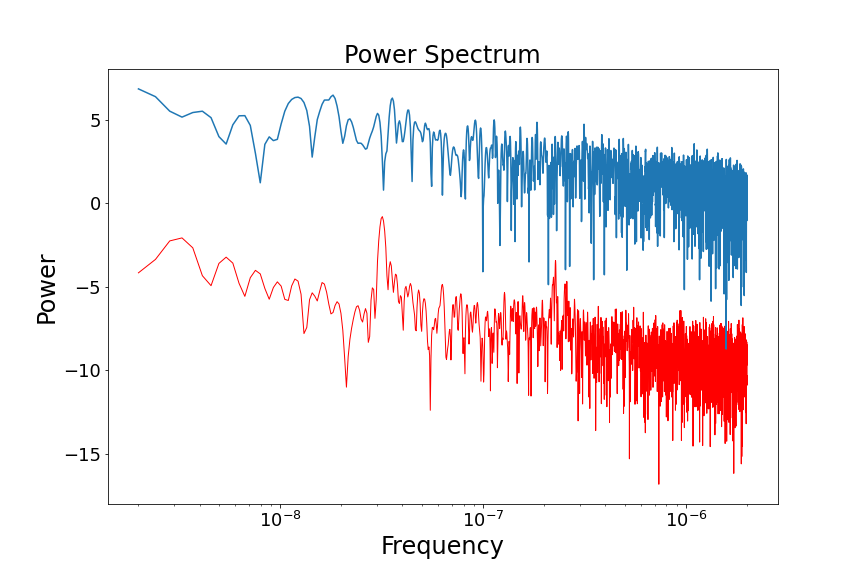}
   \caption{(left) Mkn 421 $\it{RXTE}$ ASM light curve, dwell by dwell, where the flux is in $erg \cdot cm^{-2} s^{-1} $. While the sampling is much denser, individual points have significantly lower signal-to-noise. No filtering has been applied, as is evident due to the presence of negative flux values. Mkn 421's ecliptic latitude of $~$30 degrees complicated the ASM dwell collection when the source was closest to the Sun. See \S 3.2.1.}
   \label{fig3:dbdLC}
    \caption{(right) Mkn 421 Lomb-Scargle power spectrum (blue) with an offset value of 10 has been added to the LSP for ease of comparison and window power spectrum (red) for all ASM dwells.  All three peaks noted in Figure \ref{fig2:1da_PS} are detected, and despite the lower signal-to-noise on individual points, the significance of the features is even higher. }
    \label{fig4:dbdLC}
\end{figure*}

\begin{figure}[t]
    \centering
    \includegraphics[scale=3, width=1.06\linewidth]{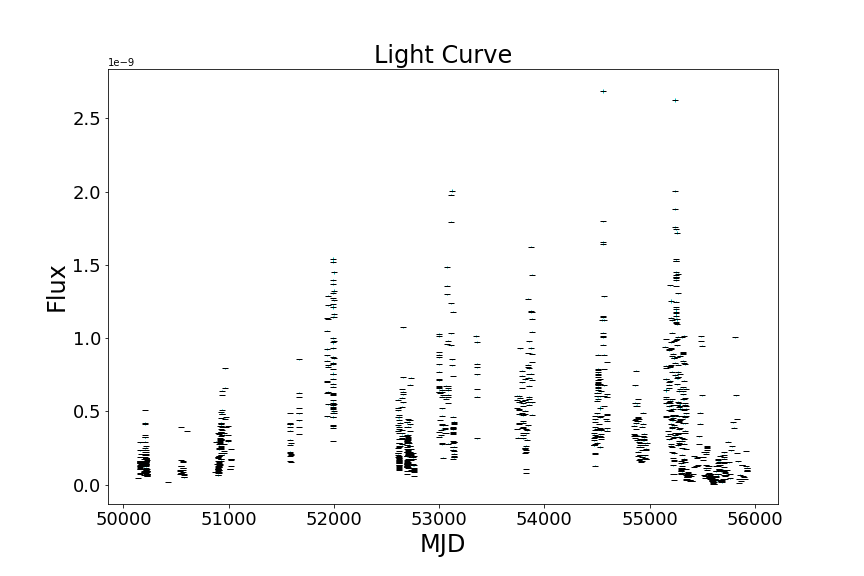}
%

       \caption{The Mkn 421 light curve of all $\it{RXTE}$ PCA observations, where the flux is in $erg \cdot cm^{-2} s^{-1} $. The signal-to-noise of individual points is much higher, but the data show fairly short campaigns, separated by $\sim 1$ year.  This is due to the target-of-opportunity nature of the observations prior to 2009. }
       \label{fig5:PCALC}

\end{figure}

\begin{figure}[t]
    \centering
    \includegraphics[scale=3, width=1.00\linewidth]{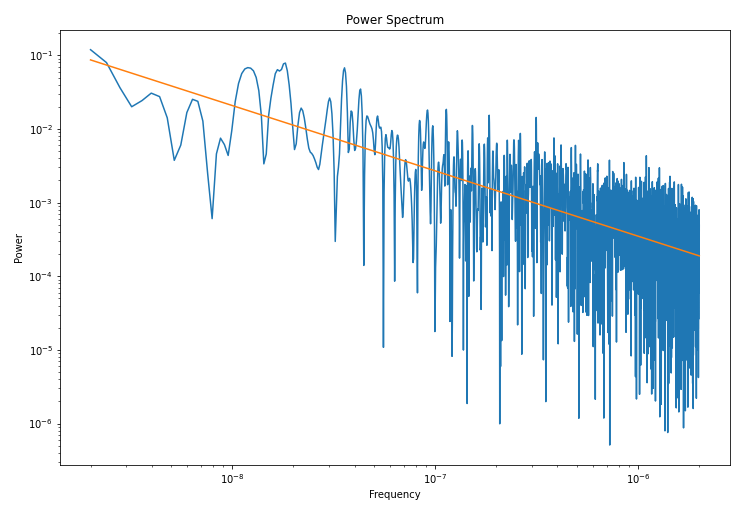}
%

       \caption{The power spectrum for Mkn 421 (blue) for the $\it{RXTE}$ ASM observations. The regression line with a slope of -0.89 used for ANOVA is shown in orange.}
       \label{fig:RegLine}
\end{figure}



Using the ASM Dwell-by-Dwell data, the obtained light curve spans the same 5837 days but contains 86935 data points with a mean spacing of 0.067 days or about 15 points a day (Figure \ref{fig3:dbdLC}). The LSP for this data (Figure \ref{fig4:dbdLC}) shows the same peak at the 2:1 resonance, corresponding to 0.001708 $\mu$Hz or a period of 650 days. This 153-$\sigma$ detection with a FAP of $\sim10^{-60}$ also shows the same peak at the 3:1 resonance, corresponding to a period of 975 days. The 325-day QPO is also present, with a significance of 90-$\sigma$ and a FAP of $\sim10^{-33}$.

A previous study done by \cite{2009ApJ...698.1207C} utilized the ASM dwell-by-dwell data for Mkn 421 from MJD 50087 to MJD 54466 (1996 January 5 $-$ 2008 July 15), manually binning the data into 20-day averages. For our analysis, Mkn 421 was bright enough that we did not encounter problems with binning. \cite{2009ApJ...698.1207C} concentrated on the power spectrum of the X-ray variability and its strength as a function of the break timescale. We found that with or without the binning, our light curves and power spectrum show the same spectral features. Additionally, cutting our data down so that it spanned the same range as \cite{2009ApJ...698.1207C}'s did not eliminate the spectral features we see in our unbinned, original runs. Our analysis benefitted from the extra data obtained after 2008 but ultimately shows that the QPO was present earlier in the mission. 

The overall power spectral density (PSD) between $2 \times 10^{-9}$ and $2 \times 10^{-6}$ Hz is well-fit by a power law, linear on the log-log plot with a slope of -0.89 (Figure \ref{fig:RegLine}). Analysis of variance (ANOVA) shows that 42.6\% of the variance in the power values is explained by the regression line (frequency dependent), while 57.3\% of the variance is explained by intrinsic random variation. When we repeat the false alarm probability (FAP) calculations using only the residuals from the regression line, the QPOs are still quite significant. The values for the FAPs calculated using the residuals can be seen in Table \ref{table:newFAP}. This table shows the values of the false alarm probabilities (FAPs) calculated using the residuals from the regression line for the 325-day, 650-day, and 975-day periods. While the values are higher than the initially calculated FAPs, the new values show that the QPOs are still quite significant.

We investigated regression fits for broken power laws. In particular, a fit to the portion above 188 days, gives a slope of -0.43 on the log-log plot. Thus the noise in the region of the reported QPO is closer to white than pink. See also \S 4.2. Using the 20-day binned data for the regression line fit as in \cite{2009ApJ...696.2170R} gave a power law slope identical to -0.89. The LSP frequency interval was shortened to above 40 days to avoid Nyquist-limited data.

\cite{2009ApJ...696.2170R} found false periods of $\sim$365 days due to instrumental effects. We see a similar phenomenon in Figure \ref{fig1:ASM_LC}. The prominent peak in the window LSP (in red) at 365 days does not seem to be influencing any of the peaks in the data LSP (in blue). Recall that the window function assumes constant flux and finds any Fourier signature due entirely to the uneven spacing of the data points. Mkn 421 has an ecliptic latitude just under 30 degrees which complicated the ASM dwell collection when the source was closest to the Sun. The false peak with a period of 1 year is commonly seen in RXTE data.

\begin{center}
\addtolength{\tabcolsep}{-3pt}
\begin{table}[H]
\centering
\begin{tabular}{||c|c|c||}
\hline
Period (days) & Z & FAP (\%) \\
\hline \hline
      325 & 16 & 0.376 \\ \hline
      650 & 16.8 & 0.157 \\
      \hline
      975 & 13 & 6.66 \\
      \hline
\hline
\end{tabular}
\caption{Values of Z and FAP calculated using the values from the regression line and ANOVA.}
\label{table:newFAP}
\end{table}
\end{center}

\subsubsection{PCA Analysis}
The PCA data cover MJD 50143-55926 (1996 March 1 - 2011 December 31), with an average spacing of 4.9 days (Figure \ref{fig5:PCALC}). There are substantial gaps. The observing strategy for Mkn 421 also had systematic bias since prior to 2009 the observations were mostly triggered as targets of opportunity, such that most observations occurred in periods when Mkn 421 was flaring. More intensive observations were undertaken between MJD 55149-55926 (2009 November 14 - 2011 December 31) with an average spacing of 2.05 days.
Figure \ref{fig10:overlappedLC} shows both the PCA and ASM light curves for this interval. We note that the flares match up for the two data sources, which strengthens the credibility of the ASM data where there are many more points but lower signal-to-noise. However, the PCA data have sparse coverage and a selection bias, while the ASM data do not. Despite these contradictions, both sets of data give similar results.
This data segment is continuous, not biased to only flaring states.
Analysis of the LSP for the PCA data shows hints of peaks at the 325-day period and its harmonics but at much lower significance. We believe that the bias toward flares, paired with substantial gaps in the observations, is more likely to produce false alarms. The LSP from the shorter light curve (2009-2011) shows a peak at 0.0429 $\mu$Hz, corresponding to a period of 270 days.
This detection is in good agreement with the 280-day period detected by \cite{2016ApJ...832...47B} and \cite{2022arXiv220413051R}.
The PCA data have a lower uncertainty (higher signal-to-noise ratio) than the ASM data. 

\begin{figure}
    \includegraphics[scale=0.3]{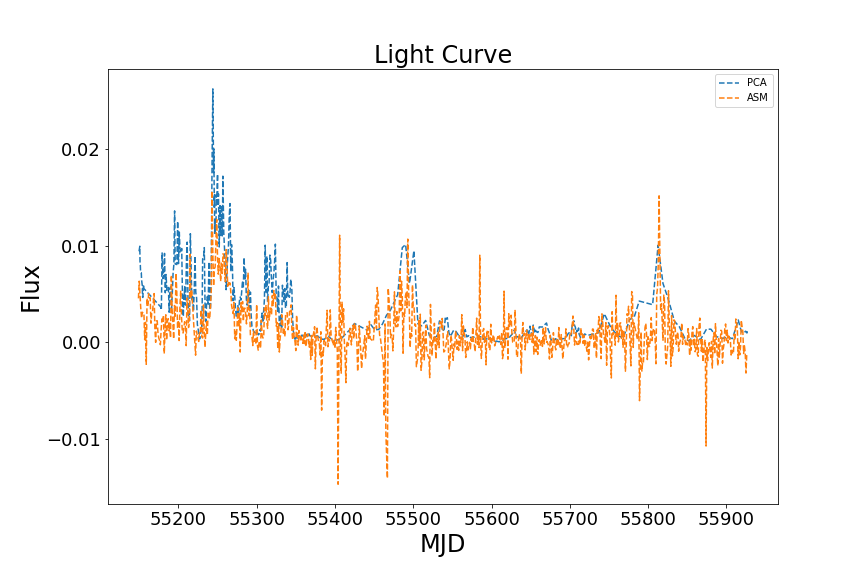}
    \caption{This figure shows the light curves for both PCA (blue) and ASM (orange) data from MJD 55200$-$55900. Flux values for both light curves are in $erg \cdot cm^{-2} s^{-1} $. The same peaks can be seen throughout the time range shown.}
    \label{fig10:overlappedLC}
\end{figure}
\vfill\eject

\section{Quality Assurance}
\subsection{Censoring of Lower Limits}
Figure \ref{fig9:markedLSP} shows the QPOs detected in our current analysis, on an expanded scale (including y-scale). We used a Gaussian fit to the conspicuous feature near 320 days to estimate its best fit value and error bar. The Gaussian fit gives a best fit period of 325 days with an uncertainty of $-7$ and $+8$ days. 
Subharmonics near 650 and 975 days also stand out. These are considerably wider than the prime QPO, and the one near 650 days shows evidence of a side lobe. A less significant peak appears at 270 days. The 325 and 270-day peaks are in good agreement with previous findings discussed in the introduction.  However, the 270-day peak does not appear significant (9.8-$\sigma$) and is not apparent in the light curve.

Outside of its original context, survival analysis can be thought of as a type of cumulative probability \citep{1992scma.conf..221F, Feigelson_Lecture}. This allows survival analysis to address censoring, or the bias introduced by excluding data. In the ASM data, left-censoring is present since we detect negative flux for our object, making it fainter than the expected limit of zero. One of the most famous estimators in survival analysis is the Kaplan-Meier estimator (KME) \citep{doi:10.1080/01621459.1958.10501452}. In astronomy, it can be used to estimate the fraction of objects above or below a certain flux level. In this case, we set the limiting flux to zero in order to address the negative flux values seen in the ASM data. This analysis showed that only two percent of all the flux values obtained by the ASM were negative. 

Traditional survival analysis, even when adapted to astronomical surveys, typically only deals with various sources, not one source with varying flux (Feigelson, private communication). We investigated other forms of survival analysis, but concluded that they were not appropriate for this analysis. We relied on our KME results and chose to leave our original FAP as the best method of verification.

 \subsection{Uneven Sampling \& Noise}Having uneven data spacing complicates some phases of the analysis, while offering the potential for enhancement in other areas. \cite{1998ApJ...504..405S} points out that binning data throws away a considerable amount of information and introduces dependency of the results on the sizes and locations of the bins. Binning means to divide the observation into equally spaced intervals, then counting photons within each bin. On the other hand, it might be argued in some cases that the portion of frequency space where information is lost may not be one where detections are expected.

The RXTE PCA data from AGN monitoring campaigns tend to be unevenly sampled in the extreme. There are gaps where no observations are proposed. For many sources (ecliptic latitude less than 30 degrees), there are yearly gaps when the source is too close to the Sun. And even where campaigns have a design mean frequency, scheduling limitations lead to a standard deviation of the spacing around the goal value. We found that the RXTE ASM data also had many of these features. Mkn 421, with an ecliptic latitude of 29.5 degrees, just missed the 30-degree PCA Sun exclusion constraint. Gaps of about 17 days are evident in the PCA light curve. While the ASM could theoretically observe at low sun angle, the Mkn 421 data also have yearly gaps. These were more complicated, averaging about 40 days, surrounded by an additional period of less than daily sampling. And the “daily averages” also use an average time (so not truly binned), thus the spacing is distributed around a mean of 1 day but with a substantial standard deviation.

{\cite{2009ApJ...698.1207C} did extensive research with the RXTE ASM database, relying on binning at 20-day or 30-day intervals (depending on source flux) the “dwell-by-dwell” data, but did not concentrate on searching for periodicities in the data. \cite{2009ApJ...696.2170R} used an approach more similar to ours, relying on the “daily averages”, while noting that there were many negative values, which were considered to be upper limits of 0 and omitted from the analysis. Their structure function (SF) analysis rendered periods in several AGN of close to 1 year (including a quoted period of 365 days for Mkn 421), which they concluded to be artifactual. See also \S 3.2.1.

Our use of the Lomb-Scargle periodogram (LSP) is guided by the suggestions in \cite{2018ApJS..236...16V}. One of our primary quality assurance tools is to compute the window function and the window transform. This is to compute the LSP using the data timestamps, but with a constant value of 1 in place of all data values. (This is the red curve often displayed with our LSP results.) Then we examine the window function for dominant features, such as daily or annual aliases or Nyquist-like limits. This is where we find that a prominent peak in the window LSP at 365 days is not seen in the data LSP. This peak is seen in a constant flux (flat) light curve based only on the uneven spacing of the light curve. So, this peak is definitely artifactual, but effectively screened out by this technique.

The Lomb-Scargle approach has been criticized because it assumes Gaussian noise, thus not appropriate for QPO/period searches in AGN light curves where red noise type variability is known to be common. Intermittent time sampling and low signal/noise of the data make it easier to mistake a phantom periodicity for a real one. As discussed in \cite{2005A&A...431..391V}, red noise in light curves with a sinusoidal appearance are produced at a rate that is not insignificant, and this rate depends strongly on the power spectrum. In this context, we note that the slope of the power spectrum is not indicative of strong red noise, but is pink, trending towards white, in the vicinity of our QPO.

Another major complication from the unevenly spaced data is in the statistical analysis of the significance of the detected peak. We implemented the False Alarm Probability (FAP) using the method of Horne \& Baliunas. The variability properties of AGN have been characterized as red noise \citep{2003MNRAS.345.1271V}, arising from a stochastic, rather than deterministic, process. Each individual light curve is deemed as merely one realization out of an ensemble of random light curves that might be generated by the underlying stochastic process. Other realizations would look different due to the statistical fluctuations inherent in any stochastic process. Two light curves will have different characteristics (e.g., mean and variance) even when they are realizations of the same process. However, data from a deterministic process (e.g., the light curve of a strictly periodic source) should be repeatable within the limits set by the measurement errors. That analysis found that the majority of the variability in the ASM lightcurve was due to the 325-day QPO, while a smaller but significant minority was due to the more quasi-random (white/pink noise dominated) variability evidenced by the power-law slope of the PSD.

\cite{2003MNRAS.345.1271V} go on to describe simulations to study properties of the variance of red noise data. Random artificial light curves were generated from power-law (AGN-like) power spectral density (PSD), then using the Timmer-Koenig algorithm \citep{1995A&A...300..707T}, random time series are generated with arbitrary broad-band PSD, to correctly account for the intrinsic scatter in the powers. Large scatter in such periodograms is an intrinsic property of stochastic processes and does not depend on the number of data points or the Poisson noise in the data. \cite{2005A&A...431..391V} introduced a relatively simple procedure to assess the significance of peaks in a periodogram when the underlying continuum noise has a power law spectrum. However, it is only applicable in the case of evenly spaced data.

We have found this to be a general problem and continue to look for other ways to assess statistical significance, in particular implementing some of the suggestions from \cite{2018ApJS..236...16V}. In Figure 6, we note that the PSD does have a linear trend on this log-log plot, matching the power-law dependence often noted in these objects. The classical Horne-Baliunas approach to the false alarm probability (FAP) assumes a flat PSD, or “white noise” spectrum. It is fair to ask if some of the significance attributed to our QPO is due to it sitting on the “up ramp”. The mean of the power values would be a straight horizontal line that typically under-predicts the power at low frequency while over-predicting it at high frequency. We computed the best fit line to the power values, and repeated the FAP calculation using residuals from that line rather than the raw values. We also did an analysis of variance (ANOVA) to determine how much of the power variance is explained by the regression line, and how much is explained by random variation. When repeating the regression fit and ANOVA calculations for our data averaged over 20-day bins, mirroring the analysis of \cite{2009ApJ...696.2170R}, we still get a slope indicative of pink or white noise.


\begin{figure}[h]
    \includegraphics[scale=0.4]{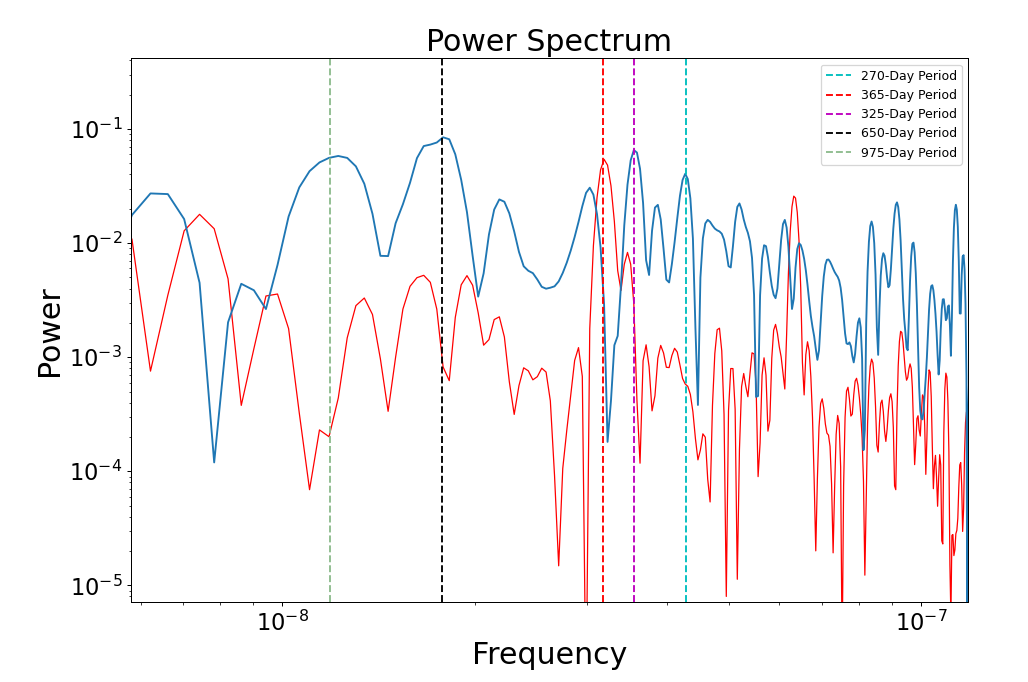}
    \caption{This figure shows the LSP from the ASM 1-day averages. Peaks of interest are labeled. See \S 3.2.1 for discussion.}
    \label{fig9:markedLSP}
\end{figure}

\section{Modulation of Oscillation}

\subsection{Keplerian Orbital Period}

The orbital timescale can be parameterized \citep{1999ApJ...514..682E, 2020ApJ...902...65S} as: 
\begin{equation}
t_{orb} \approx 0.33 M_7 \left(\frac{r}{10\,R_{S}}\right)^{1.5} \text{days}.  \label{1}
\end{equation}
$M_7$ is the mass of the black hole in units of $10^7M_{\odot}$, $r$ is the distance from the BH system barycenter, and Equation \ref{schwarschild radius} is the Schwarzschild radius.

This can be inverted to solve for $r$:
\begin{equation}
r \approx 21 \left[\frac{P(\text{days})}{M_7}\right]^{2/3} R_S.  \label{2}
\end{equation}

Using the previously scaled mass of 19.5 $M_{7}$ \citep{2004ApJ...615L...9W} as the mass of the SMBH in Mkn 421 and a period of 325 days, interpreting the QPO as a Keplerian orbital frequency gives a radial distance from the BH system center of mass of 137 $R_{S}$. 


\subsection{Lense-Thirring Precession Period}
The Lense-Thirring precession period is \citep{1998ApJ...492L..59S}:
\begin{equation}
P_{LT} = \frac{8 \pi GM}{c^3a} \left(\frac{r}{R_S}\right)^3.   \label{P_LT}
\end{equation}
Parameterizing, as before, the radius of the hot spot becomes \citep{2020ApJ...902...65S}:
\begin{equation}
\frac{r}{R_S} = \sqrt[3]{\frac{P_{LT}c^3a}{8\pi GM}} = 4.117 \;\sqrt[3]{\frac{P_{LT}(\text{days}) \;a}{M_7}}.   \label{param}
\end{equation}
So, for this QPO at 325 days, and $M_7$ = 19.5 for Mkn 421, the hot spot radius is 10.5 $R_S$ for a dimensionless spin parameter of $a \approx 1$. For a=0.1, the radial distance is roughly half this value.




\subsection{Other Possibilities}

We have concentrated on Keplerian and Lense-Thirring frequencies as these are common elements of many models. While we favor Lense-Thirring due to the smaller radius, other interpretations are possible. 


King et al. (2013) detected a QPO between 120 and 150 days over a 4-year timescale in J1359+4011 using both radio and X-ray data. While Lense-Thirring precession is considered a possibility, in this case, the timescale of the observed oscillation points to Lightman-Eardley secular instability \citep{2013MNRAS.436L.114K}. \cite{2017MNRAS.468.4992P} suggest that jet geometry varies with AGN type. Based on \textit{VLBI} data used in \textit{MOJAVE} studies, BL Lac objects are thought to have conical jets while radio jets are closer to parabolic. Mkn 421 is one of the few BL Lac objects where superluminal motion is not seen in \textit{VLBI} monitoring. \cite{2017MNRAS.468.4992P} explain this phenomenon by stating that superluminal features only take up a small piece of the entire jet cross-section due to the precession of the jet.

\section{Mechanism of Oscillation}

Simulations suggest that creating a relativistic jet requires a BH, hot inner flow, large-scale poloidal magnetic field, and large BH spin \citep{2019ARA&A..57..467B}. For these and other reasons, not all astronomical systems with accretion disks develop jets. Based on Shakura and Sunyaev’s Keplerian disk theory \citep{1973A&A....24..337S}, both the disk and jet in such states are periodic since the jet responds to the disk state. 

Looking at disk-jet coupling in AGN has proven difficult because of the long dynamic timescales associated with SMBH. \cite{2004ARA&A..42..317F} shifted their focus to binary systems since variability within these systems occurs on much shorter timescales, and they still exhibit disk-jet coupling. Based on observations of GRS 1915+105, the authors were able to create a phenomenological model that works analogously for all systems with disk-jet coupling. The model uses physical indicators to paint a quantitative picture within BHXRBs that can be generalized to include AGN. These indicators focus on the X-ray emission of the central object. They looked for prolonged hard X-ray states with a steady production of self-absorbed jets. They additionally searched for prolonged soft X-ray states with suppressed jet production and optically thin ejection events occurring during the transition between hard and soft states. The bulk velocities of transient jets associated with these events are much larger than those of steady jets. 

\cite{2004ARA&A..42..317F} state that these physical aspects give an overall picture, as follows. Jet production occurs above a certain spectral hardness, with strength correlated to the X-ray strength nonlinearly. As the spectrum of a source softens, the velocity of the jet increases in a way that is significant for the highly relativistic outflow. The source then reaches a peak in soft X-ray flux, followed by a rapid increase in jet velocity, and, finally, the jet shuts off. The rapid increase in the jet's velocity produces a shock in the preexisting jet, creating an optically thin flare with large bulk motions. A secondary flaring state in the same outburst is common. This flare is associated with further optically thin radio flares of declining strength.

A more recent review of QPO models by \cite{2019NewAR..8501524I} focuses on the relationship between a moving, characteristic radius and changes in QPO frequency. The authors favor precession as the cause of QPOs, with Lense-Thirring precession as the leading theory based on observational evidence. For this to be the case, a $5^{\circ}$ misalignment between the BH spin axis and the binary rotation axis is required. The authors present evidence for differential precession, meaning QPO frequency would increase closer to the central accreting object, similar to the increase in jet velocity described by \cite{2004ARA&A..42..317F}. While this adds to the case of Lense-Thirring precession, \cite{2019NewAR..8501524I} call for further observations to confirm their findings. 

\subsection{Disk-Jet Coupling}
The main models considered for the X-ray emission in blazars (particularly BL Lac objects) in the high state are jet based, while in the low state, blazar X-ray emission can include a significant contribution from the accretion disk, particularly for FSRQ (flat-spectrum radio quasars, \cite{2018Galax...6....1G}). In  high-state blazars, Doppler beaming effects cause emission from relativistic jets observed near our line of sight to dominate emissions over the whole spectrum. \textit{VLBI} studies have shown that the helical structure of jets can lead to a misalignment between the core and jet of blazars. The misalignment would result in rapid, large-amplitude variability as the jet moves toward or away from the line of sight thus changing the Doppler beaming factor \citep{2018Galax...6....1G, 2018NatCo...9.4599Z}. 

While QPOs are rare in intra-day variability (IDV) blazars, Mkn 421 and PKS 2247$-$131 show IDV in X-ray observations \citep{2018Galax...6....1G} as well as QPOs. The \textit{Fermi Gamma-ray Space Telescope} detected a 34.5-day QPO in PKS 2247$-$131, thought to be caused by the helical structure of the jet. Based on \textit{Very Long Baseline Array (VLBA)} monitoring of Mkn 421, \cite{2017MNRAS.469.1612L} propose the presence of two nested helical magnetic fields with opposite helicities in the jet. The variability of the rotation measure (RM) over the year-long observations implies that the jet precesses over a timescale of one year. While multiple studies concur that helical jets and their motion are the most likely cause of QPOs in BL Lacs, further observations over the whole spectrum are needed in order to be certain \citep{2017MNRAS.472.3789C, 2017MNRAS.469.1612L, 2013MNRAS.436L.114K, 2017A&A...600A.132S}.


As noted in \S 1, three recent papers
\citep{2016PASP..128g4101L,2020ApJ...905..160B,2022arXiv220413051R} have found QPOs with dominant periodicities at $\sim$300 days in Mkn 421, using data from \textit{Fermi LAT, Swift and OVRO}. In contrast \cite{2017MNRAS.472.3789C}, \cite{2020ApJS..250....1T}, and \cite{2017A&A...600A.132S} did not find QPOs in \textit{Fermi LAT} data of Mkn 421 despite their data including the same time frame. Nevertheless the claimed $\sim$300 day quasi-periodicities are consistent with the 325-day QPO we report here. While the time windows of the \textit{RXTE} (1996-2011) and \textit{Fermi} (2008-Present) observations have only modest overlap, it seems likely that this QPO may have been present in both the X-ray and gamma-ray light curves of Mkn 421 over an extended period of time, perhaps dissipating only in 2014, as claimed by \cite{2022arXiv220413051R}.  There are multiple possible mechanisms for such behavior. One, cited by  \cite{2016PASP..128g4101L}, is nonballistic helical motion of the emitting material. Another possibility is that low-frequency QPO behavior in jetted sources may be the disk response to jet instabilities, a possibility raised by \cite{2022A&A...660A..66F}. In this mechanism, the assumption, a jet instability at some distance from the disk, would lead to jet wobbling.  This instability would transmit waves both upstream and downstream.  If the waves transmitted upstream, they would connect to the disk at some point, acting like a hammer impacting the disk with a characteristic frequency. Such an instability could transfer through the disk and be modulated either at the Keplerian frequency or modulated by Lense-Thirring precession, particularly in a spinning black hole system. While the majority of the energy in the instability would advect downstream, the instability would remain near the same locus, thus leading to a relatively long-lasting and quasi-periodic disturbance.  Such a wave could arise either through pressure-driven, current-driven (kink) instabilities (e.g., \cite{2009MNRAS.394L.126M, 2011MNRAS.418L..79T}) or Kelvin-Helmholtz instabilities (\cite{2009MNRAS.394L.126M, 2015MNRAS.452.1089P}). While the \cite{2022A&A...660A..66F} model was developed for X-ray binary jets and lacks detailed MHD modeling, there is clear applicability to AGN jets as well.  At least in AGN jets it would be much less likely that the instability would have to do with jet recollimation, as in many jets that typically occurs at distances of tens to hundreds of thousands of $R_S$ from the black hole \citep{2019BAAS...51c..16P}.  However, many other such loci for instabilities are present and these modes are generally in jets.  

It should be noted that helical jet models have been linked to blazar variability in other objects, e.g., OJ 287 \citep{2013A&A...557A..28V}, CTA 102 \citep{2020A&A...642A.129S}, AO0235+164 \citep{2020ApJ...902...41W}, 0405-385 \citep{2022ApJ...931..168G}, 0716+714 \citep{2018AJ....155...31H} and 3FGL J0449.4-4350 \citep{2020PASP..132d4101Y}. They have also been invoked to explain oscillations in the polarization variability of the prototype radio galaxy M87 \citep{2011ApJ...743..119P}. In addition, helical jets have been seen on larger scales in VLBI images by \cite{2013ApJ...768...40L} for 0716+714, \cite{2016ApJ...817...96G} in the case of BL Lac, or \cite{2020A&A...641A..40V} in the case of 0836+710.   Another possibility, cited by \cite{2016PASP..128g4101L}, is nonballistic helical motion of the emitting material.  While such helical motion could be induced by a coupling between the disk and jet, it could also be excited by the same current-driven or pressure-driven kink modes described above \citep{2009MNRAS.394L.126M, 2011MNRAS.418L..79T, 2015MNRAS.452.1089P, 2018Galax...6...31A}. The main difference here is that one does not necessarily need an oscillation to transmit to the disk in order to obtain a quasi-periodic oscillation, especially if the oscillation arises at some fairly large distance from the central black hole. As above, while the wave would advect downstream, the instability that generated it would be relatively long-lived and could therefore lead to a quasi-periodic signal in the light curve.

\subsection{Accretion Disk Limit Cycle}
The accretion disk limit cycle was originally put forth to account for outbursts in dwarf novae but can also be relevant to SMBH systems \citep{1993ApJ...419..318C}. It is caused by an inherent, hysteretic relationship between vertically integrated viscous stress and surface density at a fixed radius in the accretion disk. Based on various modeling efforts \citep{1992ApJ...385...94C,1993ApJ...419..318C,1998ApJ...494..366C,2011MNRAS.414.2186J}, it is believed that the limit cycle mechanism operates in most AGN during the quasar era, solving the problem of feeding quasars at the required rates. 

In some light curves, outbursts aside from QPOs can be seen. Short-term UV variations in AGN spectra have been attributed to Lightman-Eardley instability - a phenomenon in which the accretion rate varies inversely with vertically integrated disc surface density \citep{1989MNRAS.239..289S, 1974ApJ...187L...1L}. Applying these models or assumptions to AGN can be quite difficult since they don’t have such clearly defined boundaries as in binary systems. Additionally, AGN which accrete at the Eddington rate during the high state may utilize the disk instability only as a mechanism to store matter and then trigger the instability when the surface density becomes greater than the maximum surface density when the viscosity parameter is 0.01 (disk temperature is lower) somewhere in the disk. Systems with lower supply rates may be in the thin-disk regime during both quiescence and outburst \citep{1992ApJ...385...94C}.

\cite{2011MNRAS.414.2186J} believe limit-cycle oscillations of the accretion disk may be caused by ionization instability or radiation pressure instability, with radiation pressure instability affecting galactic sources whose Eddington ratios are above 0.15 and AGN whose Eddington ratios are above 0.025. Based on their results, they tend to favor radiation and gas pressure in both galactic sources and AGN. The authors argue that radiation pressure instability applies to active galaxies since separations between outbursts are on the order of $10^6$ years for a $10^8 M_\odot$ BH, parametrization of viscosity is consistent with lack of instability in FR II radio galaxies as their Eddington ratio is less than 0.025 (lower limit for instability in active galaxies). During the hot state, the luminous core produces a radio jet, while during the cold state the radio activity ceases, and when scaled by BH mass ($10^8 M_\odot$) the $10^2$-$10^4$ year long outbursts and amplitudes are sensitive to the energy fraction deposited in the jet. This scaling gives an additional, independent argument that the intermittency in quasars on the timescales of hundreds/thousands of years is likely of a similar origin to those in microquasars. Observationally, Janiuk and Czerny push for searches for excess variability in galactic sources, outbursts that last tens to hundreds of seconds.  These would be caused by the radiation pressure instability and may not be strictly periodic or clear in periodograms \citep{2011MNRAS.414.2186J}. We believe the additional fluctuations seen in our light curves may be partially attributed to limit cycle oscillations in the disk, perhaps propagating out into the jet or in response to jet instabilities. The oscillations would not interact with Keplerian or Lense-Thirring produced oscillations. Both would contribute to the light curve in different ways.

\subsection{Future Work}
Several ongoing missions ($\it{XMM}$-$\it{Newton}$, $\it{Chandra}$, $\it{NuStar}$) perform long-stare observations that have facilitated the search for HFQPOs in AGN. Conversely, the RXTE AGN archive offers long but sparse monitoring campaigns providing the chance to find LFQPOs in AGN. We suggest that more work on additional sources in the RXTE PCA archive is warranted.

The RXTE ASM had a small collecting area compared to other ongoing missions such as Swift BAT and MAXI. Additional work with these other all-sky monitors is likely to also prove fruitful.
Studying the QPO phenomenon in AGN and BHXRB will further the understanding of both.
\acknowledgments
We thank the anonymous referee for many insightful comments that helped to improve the quality of this paper. We thank Eric Feigelson, Jacob VanderPlas, Richard Anantua, and Brandon Curd for valuable and informative discussions.

This work utilized light curves provided by the University of California, San Diego, Center for Astrophysics and Space Sciences, X-ray Group (R.E. Rothschild, A.G. Markowitz, E.S. Rivers, and B.A. McKim), obtained at \url{http://cass.ucsd.edu/~rxteagn/}. We also thank the ASM team at MIT for the light curves used in this analysis, obtained at \url{http://xte.mit.edu/ASM_lc.html}.

\bibliography{QPO}

\end{document}